\documentclass[aps, pre, twocolumn, showpacs, preprintnumbers, amsmath, amssymb, superscriptaddress]{revtex4-1}

\usepackage{mathtools}
\usepackage[sort&compress]{natbib}
\usepackage{hyperref}

\begin{document}

\title{Collective dynamics of soft active particles}

\author{Ruben van Drongelen}
\affiliation{Department of Bionanoscience, Kavli Institute of Nanoscience, Delft University of Technology, Delft, The Netherlands}
\author{Anshuman Pal}
\affiliation{Department of Physics and Astronomy, University of Pennsylvania, Philadelphia, PA, USA}
\author{Carl Goodrich}
\affiliation{Department of Physics and Astronomy, University of Pennsylvania, Philadelphia, PA, USA}
\author{Timon Idema}
\email{t.idema@tudelft.nl}
\affiliation{Department of Bionanoscience, Kavli Institute of Nanoscience, Delft University of Technology, Delft, The Netherlands}

\hypersetup{
    pdftitle={Collective dynamics of soft active particles},
    pdfauthor={Ruben van Drongelen, Anshuman Pal, Carl P. Goodrich and Timon Idema},
    pdfcreator={Ruben van Drongelen},
    breaklinks=true
}

\begin{abstract}
We present a model of soft active particles that leads to a rich array of collective behavior found also in dense biological swarms of bacteria and other unicellular organisms. Our model uses {\it only} local interactions, such as Vicsek-type nearest neighbor alignment, short-range repulsion, and a local boundary term. Changing the relative strength of these interactions leads to migrating swarms, rotating swarms and jammed swarms, as well as swarms that exhibit run-and-tumble motion, alternating between migration and either rotating or jammed states. Interestingly, although a migrating swarm moves slower than an individual particle, the diffusion constant can be up to three orders of magnitude larger, suggesting that collective motion can be highly advantageous, for example, when searching for food. 
\end{abstract}
\maketitle

\section{Introduction}
Collective migration is found throughout the living world. Examples range from shoals of fish and flocks of birds on the macroscopic level~\cite{Shaw1978,Calovi2013} to microswimmers and individual cells at the micron scale~\cite{Kessler1993,Gregoire2003,Nagano1998,Rappel1999}.  At even smaller scales within the cell, myosin motors work collectively on actin filaments to achieve long-range alignment~\cite{Butt12022010}. In such crowded environments, the simple behavior of individuals results in complex, non-trivial dynamics of the group. No individual group member can dictate the collective group behavior or even have anything close to complete information of the group's dynamics. Nonetheless, the emergent collective patterns have a huge impact on the individuals, and they often depend on them for their very survival. Therefore it is an obvious question to ask how the rules governing the behavior of each individual relate to the resulting collective behavior of the group.

In their seminal 1995 paper, Vicsek et al.~\cite{Vicsek1995} introduced a model for studying flock behavior based on a few simple rules for each individual bird. In their model, the individuals are described as oriented point particles, which exhibit self-propulsion, nearest-neighbor interactions that result in particle alignment, and noise. Many variants of the original model have been studied in the last twenty years~\cite{Vicsek2012}. Parallel to the development of the Vicsek model, much progress has also been made in the field of granular media, which studies the collective behavior of collections of large particles. In their famous 1998 Nature news and views, Liu and Nagel proposed that the observed behavior of these systems can be summarized in a phase diagram. Systems will get jammed at high densities provided both their effective temperature and the applied load are low enough, with a sharp phase transition between the jammed and unjammed state~\cite{Liu1998}. 

In recent years, several groups have started combining ideas from both fields, studying the collective behavior of finite-sized self-propelled particles. In their 2011 paper, Henkes et al.~\cite{Henkes2011} showed that for a confined system, with low self-propulsion velocities (equivalent to low load and temperature), a sharp transition can also be found between a liquid and a solid state as a function of packing density. Models without confinement often use a long-range attraction to model collective dynamics. For example Gr\' egoire et al.~\cite{Gregoire2003} combined the Vicsek model with a Lennard-Jones-like potential and found that this long-range attraction results in a cohesive flock. More recently  d'Orsogna et al.~\cite{D'Orsogna2006} and Nguyen et al.~\cite{Nguyen2012} mapped the phase space for swarms held together by a long-range Morse potential for two and three dimensions respectively. However, Rappel et al.~\cite{Rappel1999} found that long-range interaction is not a requirement for self-organization, neither in their simulations, nor in experiments (see also Wang and Kuspa~\cite{Wang1997}). 

In this paper we describe the results of our study of the collective dynamics of soft, finite sized, active particles with short-range orientational interactions, but without confinement. With no extra rules such a system would quickly fall apart. To prevent this, we assume an effective surface tension on the boundary of a cluster of particles, created by a local boundary term that directs the particles towards the cluster (i.e., particles want to move into the cluster, where the environment is usually more friendly). The density of our cluster is therefore not set by us as an adjustable parameter, but by the system resulting from a balance between its effective surface tension and the bulk modulus of the cluster. Nonetheless, we find that our system exhibits a range of different types of behavior, depending on cluster size, the particles' self-propulsion speed, and the strength of the nearest-neighbor alignment term. The two dominant types of behavior we find are collective migration and the formation of a rotational cluster with no net movement. Both are also frequently found throughout the living world. Famous examples of migrating systems are herds of mammals and aggregates of slime molds, while rotating clusters are well-known in schools of fish and the spiral of death formed by army ants. In fact, most of these systems display both types of behavior, e.g. fish switch between migration and rotation (milling)~\cite{Calovi2013,Tunstrom2013}, and depending on environmental conditions slime molds~\cite{McCann2010,Rappel1999,Vasiev1997} and bacteria~\cite{Saragosti2011,Czirok1996,Deforet2014} will migrate or rotate. For example the slime mold \textit{Dictyostelium Discoideum} (or Dicty) will collectively migrate if food is scarce, but transitions to a vortex to form a fruiting body as a last resort~\cite{Weijer2004}. Individuals in dense, biological swarms often cannot judge the volume of the swarm, but only observe their local environment. Therefore, we consider the local interaction rules we use in this work to be more realistic for describing the rules that individuals in actual swarms follow than models with long-range interactions.

\section{Method}
\subsection{Local interaction model}
We study the behavior of a two-dimensional system of self-propelling, soft, circular particles on an infinite sheet. In particular, we focus on the effects of the number of particles, the self-propulsion force and the torque that aligns the particles with each other. To prevent crystallization, the particles have different radii, drawn from a rather narrow Gaussian distribution, $G(\mu=\bar{a}, \sigma= \bar{a}/10)$, such that $\bar{a}$ is the average particle radius. The particles interact only locally. All of them experience repulsive forces when overlapping (Hookian repulsion) and Vicsek-type alignment interactions that tend to rotate their orientation to the average of that of their neighbors. Additionally, particles that are on the boundary of a particle cluster push inward, resulting in the formation of a tightly packed disordered cluster. The slight polydispersity of particle diameters, and fluctuations in the strength and direction that each particle pushes in, will lead to rearrangements and eventually large scale motion.

We apply this model to densely packed biological systems in the limit of vanishing Reynolds number. We are therefore in the regime of overdamped motion, which means that inertia is unimportant. The equations of motion for a disk in such a highly viscous fluid are given by~\cite{Landau1987}:
\begin{alignat}{2}
  \vec{F}_i &= \frac{32}{3} \eta a_i \vec{v}_i &&\equiv \alpha_i \zeta \vec{v}_i
  \intertext{and}
  T_i &= 4 \pi \eta_R a_i^2 \omega_i &&\equiv \alpha_i^2 \chi \omega_i,
\end{alignat}
with $\vec{F}_i$ and $T_i$ the net force and torque acting on particle $i$, $a_i$ the particle radius, and $\alpha_i = a_i/\bar{a}$ the normalized radius. The effective translational and rotational viscosity are $\eta$ and $\eta_R$, respectively, and $\vec{v}_i$ and $\omega_i$ are the linear and angular velocity of the particle. To simplify our expressions, we define the rescaled viscosities $\zeta = (32/3)\eta$ and $\chi = 4 \pi \eta_R$.

We denote the position of particle $i$ by $\vec{x}_i$ and its orientation by $\hat{\psi}_i$. Particles are considered neighbors for the purpose of the orientation interaction if their centers are less than $2.7 \bar{a}$ apart. With this cut-off distance, two touching particles with radius $a_i = 1.3 \bar{a}$ will still be considered neighbors, but two small particles ($a_i = 0.7 \bar{a}$) separated by a third small particle will not. Because the spread in the radius is $\sigma = \bar{a}/10$, the probability of finding even larger or smaller particles together is negligible. 

\begin{figure}[ht]
  \centering
  \includegraphics[width=0.9\columnwidth]{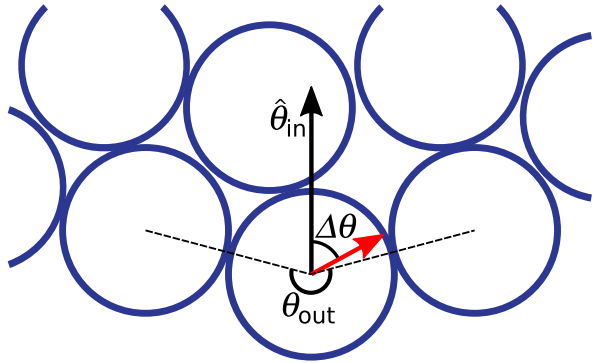}
  \caption{Visualisation of the boundary rule. The particle at the bottom finds no neighbors over an angle $\theta_\text{out} > \pi$. It therefore exerts a torque in order to align its orientation $\hat{\psi}_i$ (small arrow) to the exterior bisector of $\theta_\text{out}$, which is denoted $\hat{\theta}_\text{in}$ (long arrow). The torque it exerts scales linearly with $\Delta \theta$, the angle between these two vectors. Simultaneously, the particle exerts an additional force in the direction of orientation, proportional to $\theta_\text{out} - \pi$.}
  \label{fig:inwards}
\end{figure}

Instead of an attraction or geometrical confinement, our model uses a local boundary term to prevent systems from falling apart. An individual looks at the positions of its neighbors to determine its position within the cluster. If particle $i$ has no neighbors over an angle $\theta_{\text{out},i} \geq \pi$ we consider it to be on the boundary of the cluster and it exerts an additional torque and force (see Fig.~$\ref{fig:inwards}$ for relevant quantities). Particles with only one or two neighbours automatically satisfy this criterion. Let $\mathcal{N}_i$ denote the set of neighbors of particle $i$. The net force and torque on the particle are then given by
\begin{align}
\label{modelforce}
\vec{F}_i &= \vec{F}_{i,\text{self-propulsion}} + \vec{F}_{i,\text{boundary}} + \vec{F}_{i,\text{repulsion}} \nonumber \\
          &= \left[F_\text{self} + \left( \theta_{\text{out},i} - \pi \right) F_\text{in} \Theta( \theta_{\text{out},i} - \pi ) \right] \hat{\psi}_i - k \sum_{j \in \mathcal{N}_i} \vec{d}_{ij}, \\
\label{modeltorque}
T_i &= T_{i,\text{boundary}} + T_{i,\text{noise}} + T_{i,\text{align}} \nonumber \\
    &= T_\text{in} \Delta\theta_i \cdot \Theta( \theta_{\text{out},i} - \pi ) + T_\text{noise} \xi_i + \frac{T_\text{align}}{\left| \mathcal{N}_i \right|} \sum_{j \in \mathcal{N}_i} \Delta\psi_{ij},
\end{align}
where $\Theta(\theta)$ is the Heaviside step function. In Eq.~(\ref{modelforce}), the first two terms of the force are the self-propulsion and the boundary force, which act in the direction of orientation~$\hat{\psi}_i$. The strength of these interactions is set by $F_\text{self}$ and $F_\text{in}$ respectively. The last force term is the repulsion between overlapping particles $i$ and $j$, where the amount of overlap is given by $|\vec{d}_{ij}|$ (which of course is zero for non-overlapping particles). The strength of the repulsion force is set by the spring constant~$k$. The first term of the torque in Eq.~(\ref{modeltorque}) turns particles on the boundary inwards. The torque is proportional to a parameter $T_\text{in}$ times the angle between the orientation $\hat{\psi}_i$ and the exterior bisector of $\theta_{\text{out},i}$. The second term is responsible for the orientational noise a particle experiences. We pick $\xi_i$ randomly from $\left\{ -1, 1 \right\}$ each timestep creating a torque of magnitude $T_\text{noise}$. The final term of the torque aligns particles to the average orientation of their neighbors, where $T_\text{align}$ is the interaction strength, $\left| \mathcal{N}_i \right|$ is the number of neighbors, and $\Delta \psi_{ij}$ is the mismatch in orientation between particles $i$ and $j$. The alignment is the only interaction which acts between particles (apart from the passive repulsion) and is therefore ultimately responsible for collectivity in Vicsek-type models. Note that, in analogy with the Vicsek model, we only include noise on the torque and not on the force. With this noise term, the motion of a single particle becomes a random walk; a single noise term is thus sufficient to introduce an element of randomness in each particle's motion, and additional noise terms do not qualitatively change our results. Eliminating the noise on the torque (and hence the orientations) on the other hand does have a strong effect, as this noise term is required to obtain the rich behavior we observe.

\subsection{Simulations}
To characterize the behavior of our system, we introduce dimensionless scaling parameters that represent the strengths of the various interactions. We define $\tau = \zeta/k$ as the characteristic timescale for two overlapping particles to separate due to their repulsive interaction. For any other interaction $X$ we define a scaling parameter $\lambda_X = \tau / \tau_X$, where $\tau_X$ is the characteristic timescale of interaction $X$. The characteristic timescales and scaling parameters for all interactions in our model system are given in Table~\ref{tab:interactions}.
\begin{table}[htp]
  \begin{tabular}{|p{0.25\columnwidth}|l|p{0.35\columnwidth}|}
  \hline
  Interaction & Timescale & Dimensionless scaling parameter \\
  \hline
  Repulsion
    & $\tau = \zeta / k$ 
    & - \\
  \hline
  Alignment 
    & $\tau_\text{align} = \chi / T_\text{align}$ 
    & $\lambda_a = \zeta T_\text{align}/ k \chi$ \\
  \hline
  Noise
    & $\tau_\text{noise} = 2 \chi^2 / T_\text{noise}^2 \Delta t$ 
    & $\lambda_n = \zeta T_\text{noise}^2 \Delta t / 2k \chi^2$ \\
  \hline
  Inward force
    & $\tau_{F_\text{in}} = \zeta \bar{a} / F_\text{in}$
    & $\lambda_{F_\text{in}} = F_\text{in} / k \bar{a}$ \\
  \hline
  Inward torque
    & $\tau_{T_\text{in}} = \chi / T_\text{in}$
    & $\lambda_{T_\text{in}} = \zeta T_\text{in} / k \chi$ \\
  \hline
  Active force / Self-propulsion
    & -
    & $\lambda_s = F_\text{self} / k \bar{a}$ \\
  \hline
  \end{tabular}
  \caption{List of characteristic timescales and scaling parameters for all interactions in our simulations. For the inward force, we used the approximation that $2\arctan\left(\frac{x}{2\bar{a}}\right) \approx \frac{x}{\bar{a}}$. Since self-propulsion is an active process, it does not have a characteristic relaxation timescale.}
  \label{tab:interactions}
\end{table}

In Table~\ref{tab:interactions} we denote the duration of one Monte Carlo simulation step by $\Delta t$. The self-propulsion has no characteristic timescale as it corresponds to an external rather than a restoring force. To arrive at a dimensionless parameter describing the strength of the self-propulsion, we define $\lambda_s \equiv F_\text{self} / (k \bar{a})$, in analogy with the inward force exerted by boundary particles. We choose our unit of length by setting the average radius of the particles to unity, i.e. $\bar{a} = 1$. We set the force scale by choosing the repulsion coefficient $k = 1$. We fix our unit of time by letting the characteristic timescale of repulsion be unity: $\tau = \zeta/k = 1$. Furthermore, we may set $\chi = 1$, since we can set the strength of all torques individually~\footnote{We need to set $\chi$ because Stokes' paradox does not allow us to relate the translational and rotational viscosities $\eta$ and $\eta_R$ ~\cite{Landau1987}.}. 

A direct consequence of the nearest neighbor alignment and the presence of a non-negligible inwards torque, is that there will be some alignment mismatches, or defects, inside the cluster. Topology dictates that a simply connected cluster must have at least one such defect. We find that these defects act as organizing centers for the particles. Therefore, multiple defects either quickly coalesce or cause the cluster to break up into smaller clusters, each with its own defect. To ensure that no more than one defect will exist, we initialize our simulations by placing the particles on a square lattice in a rectangular shape with a width of 10 particles, with a small deviation from the exact lattice points. Furthermore, we set the initial direction along the long edge of the rectangle, with a deviation up to $\pi/4$ radians. We then run our model for $10^8$ steps for a total number of $N=1$, $N=100$, $N=200$, $N=400$, $N=800$, $N = 1000$ or $N=1600$ particles with alignment coefficients $0.1 \leq \lambda_a \leq 1$ and $0.04 \leq \lambda_s \leq 0.08$. We keep the other interactions constant for all simulations, i.e., $\lambda_n = 0.03$, $\lambda_{F_\text{in}} = 0.3$ and $\lambda_{T_\text{in}} = 3$. By choosing these values we ensure that the noise never exceeds the alignment, the boundary force is small compared to the repulsion, and particles on the boundary will turn inwards for even the largest value of the alignment parameter $\lambda_a$.

\begin{figure}[htp]
  \centering
    \includegraphics[width=\columnwidth]{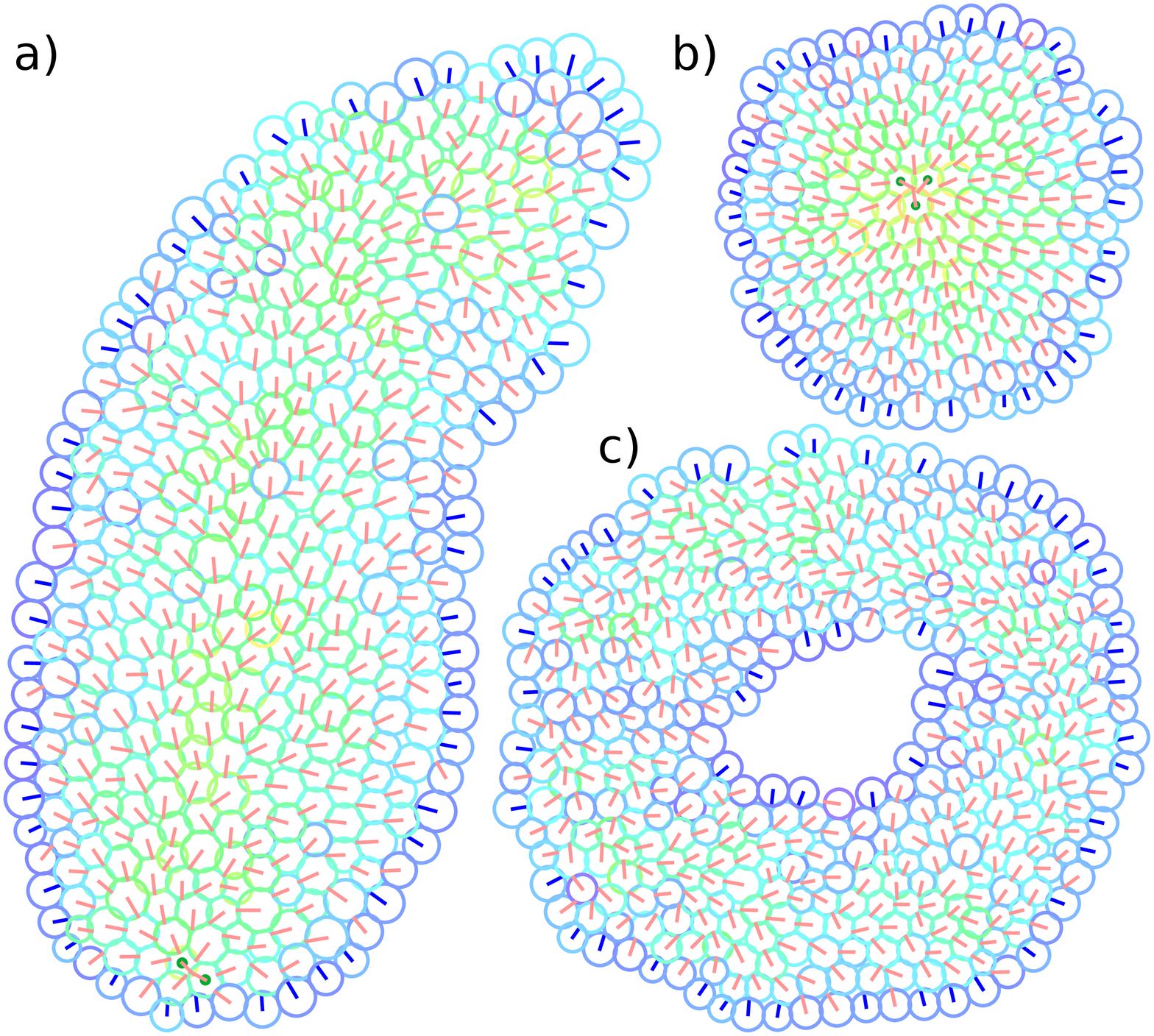}
    \caption{Typical snapshots for a) migrating $N=400$, b) jammed $N=200$ and c) rotating $N=400$ clusters. The color code blue-green-yellow-red indicates the degree of overlap with neighboring particle in increasing order. Each particle's orientation is shown by a line originating from the particle's center. This line is red for particles in bulk or blue for particles on the boundary exerting additional force and torque.}
  \label{fig:colonies}
\end{figure}

We find four main types of behavior. The cluster of particles can remain simply connected and migrate either randomly (type 1: migrating; see Fig.~\ref{fig:colonies}a), or ballistically without internal rearrangements (type 2: jammed; see Fig.~\ref{fig:colonies}b). Alternatively, the cluster can change its topology by either breaking apart (type 3: breakup; not shown) or transforming into a doughnut shape with a hole in the middle (type 4: rotating; Fig.~\ref{fig:colonies}c). We can distinguish these types of behavior by looking at the cluster's orientational order parameter, defined as
\begin{equation}
  \phi \equiv \frac{1}{N} \displaystyle \left| \sum_{i=0}^N \hat{\psi}_i \right|.
  \label{eq:order}
\end{equation}
A high value of the order parameter tells us that the cluster has a net migration direction. A low value of the order parameter means that the individual particle orientations effectively cancel and the cluster is either jammed or rotating in place. The latter two types of motion are easily distinguishable visually.

We save the average location of the particles, the location of the defect and the value of the order parameter every $128$ steps. If the cluster breaks up, the order parameter will show a slight drop. We can verify the break up by plotting the location of the defect. If we find multiple defects, or we find that the average position does not follow the defect like a trailer follows a car, we conclude that the topology of the cluster has changed. We find the diffusion coefficient $D$ of the cluster from the velocity autocorrelation function of the average location of the particles. To do so reliably, we discard the first $10^6$ simulation steps to eliminate the effects of the transition from the initial configuration to the shape the cluster naturally takes when migrating.

\section{Results and discussion}
\subsection{The order parameter characterizes behavior}
We ran 10 simulations each for cluster sizes $N = 100$, $N = 200$, $N = 400$, $N = 800$, $N = 1000$ and $N = 1600$, seven values of the alignment strength $\lambda_a$, and five values of the self-propulsion, $\lambda_s$. We found rich state behavior. A lack of alignment resulted in the cluster breaking up, whereas very strong alignment in combination with little activity resulted in a jammed system. In the jammed state, all particles are oriented towards the center of the cluster and there are very few rearrangements (see Fig.~\ref{fig:colonies}b and Movie 1 of~\footnote{See Supplemental Material at [URL will be inserted by publisher] for movies of the various types of behavior displayed by the cluster}). For intermediate values of the alignment strength, the cluster forms an elongated structure (see Fig.~\ref{fig:colonies}a and Movie 2 of~\cite{Note2}). This `slug' has its orientational defect close to the leading edge, dictating more or less the direction of motion. The exact location of the defect is subject to random fluctuations, because of the noise on the particle orientations. Hence, the movement of the slug is a random walk. Finally, for high activity or weak alignment, the cluster eventually folds onto itself, creating a vortex state (see Fig.~\ref{fig:colonies}c and Movie 3 of~\cite{Note2}). In a vortex, all the particles revolve around a common center such that the net movement is canceled out. The defect is removed by the creation of a hole in the middle.

\begin{figure}[htp]
  \centering
    \includegraphics[width=\columnwidth]{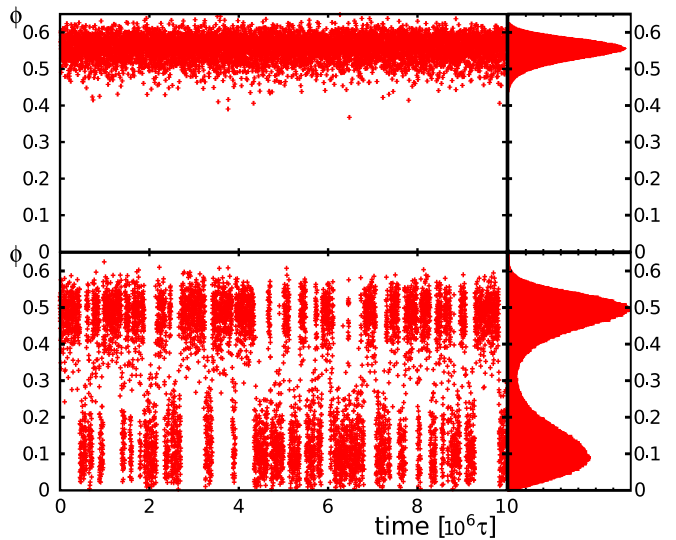}
    \caption{Evolution of the order parameter $\phi$ for $N=400$, $\lambda_s = 0.07$ and $\lambda_a = 0.67$ (top), and $N=400$, $\lambda_s = 0.07$ and $\lambda_a = 0.30$ (bottom) over time in units of $\tau$. The histogram in the top panel has one peak at $\phi \sim 0.55$, which indicates that the cluster is in the migration state. The global behavior of the cluster in the bottom panel constantly switches between migration (with $\phi \sim 0.5$) and rotation (with $\phi \sim 0.1$). The right hand panel shows the associated histogram with a bimodal distribution that represents two distinct types of behavior.}
    \label{fig:order}
\end{figure}

We can distinguish between the different states using the order parameter (Eq.~\ref{eq:order}). Fig.~\ref{fig:order} displays two examples of the evolution of the order parameter during the simulation, as well as their histograms. For the migration state (top panel), we find only one peak in the histogram. The migration state is characterized by an order parameter $\phi > 0.25$. In the rotation state, the histogram also has a single peak, but at lower values of the order parameter, $\phi < 0.15$. The jammed state can have a peak at any value of $\phi$, depending on the configuration it got stuck in. Since jammed states follow straight or circular paths (in contrast to the random walk of migrating clusters and stationary position of rotating clusters), distinguishing between jammed, rotating and migrating clusters is easy. At the boundaries between migrating and jamming, and between migrating and rotating, we find chimeric or mixed states that perform a kind of run-and-tumble motion (see Movie 4 of~\cite{Note2}). The associated histogram of the order parameter $\phi$ has two peaks, as shown in the bottom panel of Fig.~\ref{fig:order}.

\begin{figure}[htp]
  \centering 
  \includegraphics[width=\columnwidth]{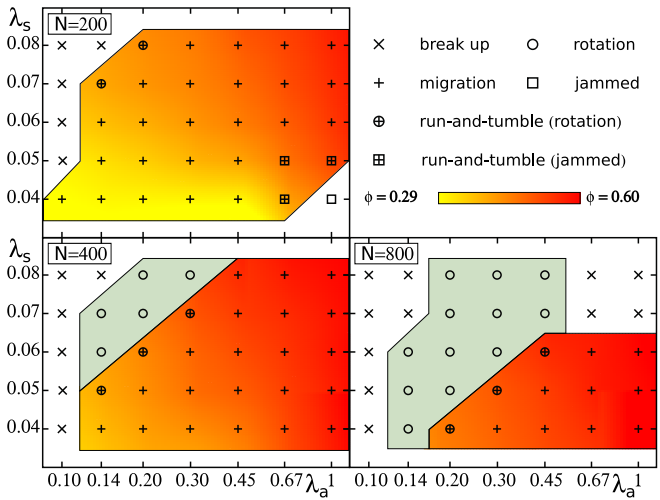}
  \caption{State diagrams of the global behavior for $200$, $400$ and $800$ particles with varying activity per particle $\lambda_s$, and alignment strength $\lambda_a$. The cluster may break up $(\times)$, migrate $(+)$, form a vortex ({\footnotesize{$\bigcirc$}}) or jam with all particles facing the center of the cluster ($\square$). Clusters may also perform run-and-tumble motion, a chimeric state mixing two types of behavior. These mixed states are denoted with the chimeric symbols $\oplus$ and $\boxplus$ for migration with rotation and jamming, respectively. The degree of alignment is measured by the order parameter $\phi$ for migrating clusters. Yellow corresponds to low values of $\phi$, red to high values of $\phi$. The green area corresponds to purely rotating clusters. Lines are guides to the eye.}
\label{fig:state_diagram}
\end{figure}

\subsection{State diagram}
We have captured the various types of behavior in state diagrams (Fig.~\ref{fig:state_diagram}). For $N=200$ particles, most of the state diagram is occupied by migrating colonies ($+$). For low activity, the system jams like passive granular matter at high density ($\square$). A strong alignment contributes positively towards jamming by preventing rearrangements (see Fig.~\ref{fig:colonies}b). For very weak alignment, the cluster is disordered and falls apart ($\times$). By increasing the number of particles these states shift towards the bottom right, making room for another state between break up and migration. At high activity and weak alignment, a migrating cluster is likely to fold onto itself. This creates a vortex state ({\footnotesize{$\bigcirc$}}) where all particles circle around the topological defect, which can even be resolved by a gap in the middle (see Fig.~\ref{fig:colonies}c). Increasing the number of particles further ($N = 800$) continues the trend of shifting towards the bottom right. The break ups in the top right corner (high alignment and high self-propulsion) are caused by particles falling off the tail of a migrating cluster due to its strongly elongated structure. Boundary particles literally pinch off small pieces of the 3-4 particle wide tail until the main cluster reaches a stable size.

The most interesting points in the state diagrams are the points between pure migration and rotation, and between migration and jamming ($\oplus$ and $\boxplus$ respectively). We observed chimeric states where both types of global behavior are present. The resulting motion is a run-and-tumble. When the order parameter has a high value the cluster migrates. During migration the defect can move towards the middle of a cluster due to the noise on the individual orientations and enter the jammed state or the rotation state with a low order parameter. The same noise is responsible for undoing this process, and allow the cluster to resume migration, in a direction independent of the direction before it went into the state of low order (see Movie 4 of~\cite{Note2}). The bottom panel of Fig.~\ref{fig:order} shows the evolution of the order parameter and the corresponding histogram for a chimeric state between migration and rotation. The time between transitions increased dramatically when we increased the value of the alignment parameter. For $N=800$ particles at $\lambda_a=0.45$, the typical time the cluster spends in one of the two states was of the same order as our default simulation length ($10^7 \tau$). The states last long because the transitions happen when the defect has moved from the boundary to the center by random fluctuations. A high alignment parameter limits the mobility of the defect within the cluster.

Note that Fig.~\ref{fig:state_diagram} shows that the global behavior changes with the number of particles. For example, an aggregating cluster can change from migration to run-and-tumble to pure rotation by collecting more particles on its way. No particle is aware of the size of the cluster. Consequently, even though the local interactions remain the same, the global behavior can change dramatically. 

\begin{figure}[ht]
  \centering
  \includegraphics[width=\columnwidth]{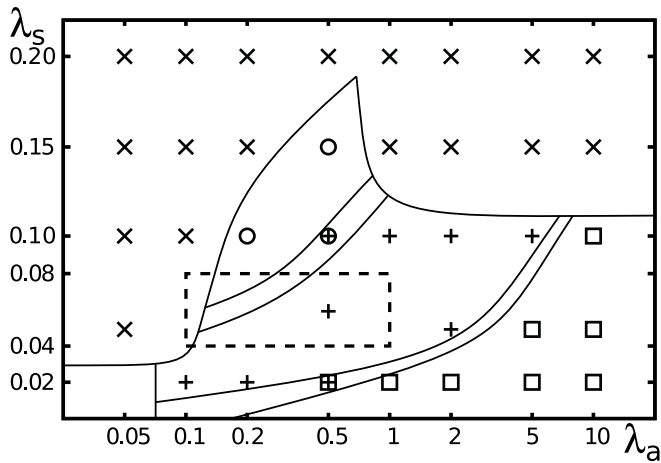}
  \caption{Zoomed out version of the $N=400$ state diagram in Fig.~\ref{fig:state_diagram}. The dashed rectangle corresponds to the region shown in Fig.~\ref{fig:state_diagram}. Boundaries between states generally shift towards the bottom right corner with increasing particle number. Chimeric states form a bridge between rotation and migration, and between migration and jammed.}
  \label{fig:extreme}
\end{figure}

We also constructed state diagrams for $N=100$, $N=1000$ and $N=1600$ particles. The $N=100$ diagram showed many signs of finite size effects. Clusters of only $100$ particles have a large number of particles on the boundary. Statistical fluctuations on the order parameter became so large that characterizing the states was far from trivial. Unsurprisingly, the state diagram for $N=1000$ looks very similar to $N=800$. Also the state diagram of $N=1600$ shows no surprises with only break ups and vortex states. Furthermore, we did some simulations with extreme values for the alignment and self-propulsion parameters for $N=400$ in order to see where the transition lines are and how they move when changing the size of the cluster (Fig.~\ref{fig:extreme}). We retrieved the jammed state for a high value of the alignment or a low self-propulsion. We also find chimeric states at higher values of $\lambda_a$ and $\lambda_s$, which suggests that transitions are smooth and we can easily tune the parameters such that the amount of time the cluster spends in either state is equal. Finally, there is a small unlabeled region with low activity and low alignment, where the self-propulsion is too low to tear the boundary apart. At the same time, the alignment is too weak to overcome the noise, such that the particles rotate randomly while hardly moving.

\subsection{Migrating collectively boosts the diffusion constant}
A large fraction of our state diagrams is taken up by migrating clusters ($+$). These clusters perform random walks on the infinite plane. The movement of the cluster is guided by the location of the defect since most particles are pointing towards it. However, the clusters are very dynamic, and particles take turns being close to the defect. In the bulk, particles move towards the defect. At the defect, the pressure is higher than the surface tension provided by the boundary particles. This allows particles from the bulk to escape into the boundary at the leading edge. At the boundary, particles move towards the trailing end of the cluster since they are now pointing in a different direction than the cluster's net motion. Once they are close to the trailing end, the pressure in the bulk is lower and the particles can penetrate in to repeat the cycle (see Movie 2 of~\cite{Note2}). 

\begin{figure}[ht]
  \centering
  \includegraphics[width=\columnwidth]{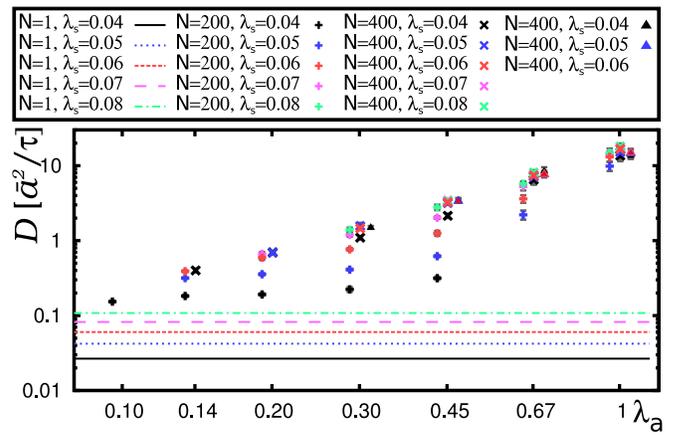}
  \caption{Diffusion constants of the center of mass of $N$ particles, in units of $\bar{a}^2/\tau$, for different values of alignment strength $\lambda_a$. Lines correspond to $N=1$, pluses to $N=200$, crosses to $N=400$ and triangles to $N=800$ (plotted next to each other for clarity). Colors correspond to different values of the self-propulsion strength $\lambda_s$: black (solid line) $\lambda_s=0.04$, blue (dotted line) $\lambda_s=0.05$, red (short dashed line) $\lambda_s=0.06$, pink (long dashed line) $\lambda_s=0.07$, light green (dot-dashed line) $\lambda_s=0.08$. Collective migration can increase the diffusion constant by up to three orders of magnitude.}
  \label{fig:D_align}
\end{figure}

We calculated the diffusion constant $D$ of these migrating clusters using their velocity auto correlation function and their mean square displacement. We found a significant increase in the diffusion constant for clusters compared to single particles. In Fig.~\ref{fig:D_align} we plot the diffusion constant for migrating clusters with $N=1$ (lines), $N=200$ (pluses), $N=400$ (crosses) and $N=800$ (triangles). Different colors represent different values of the strength of the self-propulsion force. We see that the diffusion constant is larger (up to three orders of magnitude for strong local alignment) than for a single particle. Hence, organisms in swarms may follow similar rules as described in our model to quickly explore large regions when looking for resources.

For single particles we verified that the diffusion constant scales quadratically with the self-propulsion strength, $D \propto \lambda_s^2$, whereas for large clusters the individual velocities hardly affect the diffusion constant. Similarly, we find that the persistence length $\ell_p$ increases with $\lambda_s$ for small ($N=1$ and $N=200$) clusters, but slightly decreases with $\lambda_s$ for larger ($N=400$ and $N=800$) clusters. In small clusters an increased activity causes the path length between turns (and thus the persistence length) to increase. In contrast, larger clusters will turn more quickly when they become more active. A possible explanation may be that a higher amount of activity pushes the defect forward, closer to the boundary. Consequently, the cluster changes its shape and becomes longer and narrower when $\lambda_s$ increases. With fewer particles at the tip, displacements of the defect are less damped, resulting in more and sharper turns. Therefore, both the persistence length decreases, and the likelihood of the cluster entering the rotation state increases.

To appreciate the relation between the diffusion constant and the alignment strength, we work out the Green-Kubo relation in two dimensions~\cite{Visscher1973}. Let $\vec{v}_c$ be the velocity of the center of mass of the cluster, $\ell_p$ the persistence length of its path and $\theta(t)$ the angle between $\vec{v}_c(0)$ and $\vec{v}_c(t)$. The diffusion constant is then given by
\begin{align}
  D &= \frac{1}{2} \int_0^\infty \left\langle \vec{v}_c(0) \cdot \vec{v}_c(t) \right\rangle \text{d}t = \frac{v_c^2}{2} \int_0^\infty \left\langle\cos\left(\theta(t)\right)\right\rangle \text{d}t \nonumber \\
    &= \frac{v_c^2}{2} \int_0^\infty e^{-\frac{v_c t}{\ell_p(\lambda_s, \lambda_a)}} \text{d}t = \frac{1}{2} v_c \ell_p(\lambda_s, \lambda_a), \label{eq:diff_exact}
\end{align}
where we approximated that $\ell(t) \approx v_c t$, i.e., the length of the path traveled by the average position of the cluster can be approximated by the product of the average velocity $v_c$ and the time interval. We see that a more persistent trajectory leads to a higher diffusion coefficient. The alignment counters the noise that is responsible for diffusion in the first place. The increased persistence makes clusters diffuse faster, even though the net speed of the cluster is less than the self-propulsion speed of a single particle - that is, clusters are slower than individuals because the particles are not all perfectly aligned. In fact, we can use the order parameter $\phi$ to derive the velocity of the cluster.
\begin{align}
\label{clustervelocity}
  \vec{v}_c &= \frac{1}{N} \displaystyle \sum_i^N \vec{v}_i = \frac{1}{N} \sum \frac{\vec{F}_i}{\alpha_i \zeta} \nonumber \\
    & \begin{aligned}= \frac{\bar{a}}{N \tau} &\left( \sum_i^N \frac{\lambda_s \hat{\psi}_i}{\alpha_i} + \sum_{i, j\neq i}^N \frac{\vec{d}_{ij}}{\alpha_i \bar{a}} \right. \\
    &\quad + \left. \smashoperator{\sum_{i \in \text{boundary}}} \frac{\lambda_{F_\text{in}} \hat{\psi}_i \left( \theta_{\text{out},i} - \pi \right)}{\alpha_i} \right). \end{aligned}
\end{align}
The second term in Eq.~\ref{clustervelocity} drops out since $d_{ij} = - d_{ji}$, if we neglect the effects of polydispersity on the velocity by setting $\alpha_i = 1$. The last term will also be small since the inward force by particles on opposing sides of the cluster tend to cancel out. We thus arrive at
\begin{equation}
\label{eq:v_c_exact}
\left| \vec{v}_c \right| = \frac{\bar{a} \lambda_s \phi}{\tau}.
\end{equation}
Because a higher value of the alignment strength~$\lambda_a$ results in an increase of the order parameter, both the persistence length $\ell_p$ and the cluster velocity $\vec{v}_c$ increase with $\lambda_a$.

\begin{figure}[ht]
  \centering
  \includegraphics[width=\columnwidth]{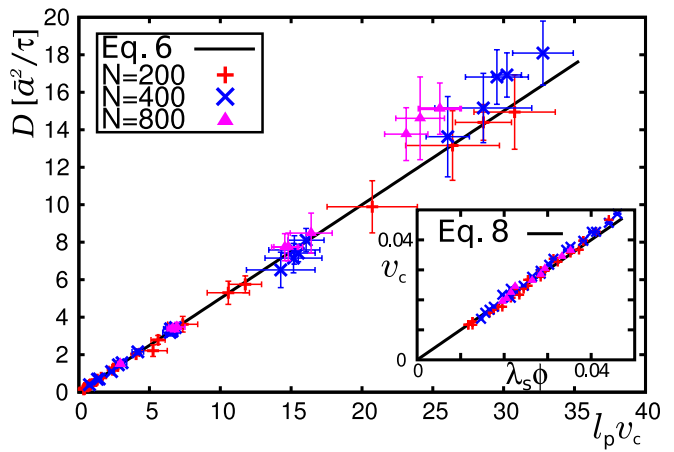}
  \caption{The diffusion coefficient of migrating clusters, in units of $\bar{a}^2/\tau$, as a function of $\ell_p v_c$, for $N=200$ (red pluses), $N=400$ (blue crosses) and $N=800$ (magenta triangles) particles. The black line is the exact result from Eq.~\ref{eq:diff_exact}. Inset: The speed of migrating clusters with self-propulsion strength $\lambda_s$ times order parameter $\phi$. The black line is the exact relation from Eq.~\ref{eq:v_c_exact}.}
  \label{fig:analytic}
\end{figure}

To verify that the assumptions made in deriving Eqs.~\ref{eq:diff_exact} and~\ref{eq:v_c_exact} are justified, we plot both relations in Fig.~\ref{fig:analytic}, together with our simulation data. The assumptions are that the speed of the cluster $\left| \vec{v}_c \right|$ is constant in time and that the polydispersity of the particles has little effect on the magnitude of the forces. The polydispersity merely serves as a way to prevent crystallization. From our simulations, we find that the diffusion constant $D$ depends linearly on $\ell_p v_c$ with slope $\frac{1}{2}$ for all cluster sizes, consistent with Eq.~\ref{eq:diff_exact}. Towards higher values of $\ell_p v_c$, determining the persistence length and diffusion constant becomes harder as the simulation is finite. The inset of Fig.~\ref{fig:analytic} shows that Eq.~\ref{eq:v_c_exact} holds for all cluster sizes.

\subsection{Migrating and rotating states in biology}

We have shown that simple, and from the perspective of the individual, sensible rules on local scales lead to various types of behavior that are relevant for biological organisms. The fast collective migration state for example is useful when exploring large areas for food. This mechanism is used by both amoebae~\cite{Weijer2004} and bacteria~\cite{Deforet2014}. The rotation state is often observed as the onset of the formation of a fruiting body which are formed by among others, the amoeba Dictyostelium Discoideum~\cite{Weijer2004} and Myxobacteria~\cite{Shapiro1988}. Furthermore we found a state where the system can switch between collective migration and stationary rotation. The ratio of time spent in one of these two types of behavior is quite sensitive to changes in activity or alignment strength. This sensitivity allows the system to easily switch between migration and rotation when the environment changes.

\section{Conclusion}
We have shown that finite, stable clusters of self-propelled soft particles can be formed with only local rules. The boundary rule that we introduced creates an effective surface tension for our clusters, which prevents their breakup. The rule also dictates the presence of at least one defect in each cluster. We found that these defects dominate the clusters' global dynamics. Elongated, slug-like migrating clusters exhibit enhanced motility with a diffusion constant that can be up to three orders of magnitude higher than that of an individual particle. The high diffusion constant demonstrates how clustering can be a good strategy for organisms in environments that are hostile or scarce in food. For larger clusters, there is a spontaneous transition to a topologically and dynamically different state: a doughnut-shaped rotating cluster with no net movement. Clusters can be brought from the moving to the stationary rotating state simply by growing in size, without the need for an additional decision mechanism. 

The vortex and migration state, and the migration and jammed state are separated by chimeric states where both types of behavior are present. The average time the cluster spends in each state can be controlled by changing the strength of alignment between particles or the self-propulsion of the particles. Therefore, in contrast to the jamming transition, which occurs at a critical density, we find no single critical value for the strength of alignment nor for the self-propulsion. Instead, we find a gradual transition where both states (migration and rotation/jammed state) coexist. With our model the density of our clusters cannot be set a priori, so we could not verify the observation by Henkes et al.~\cite{Henkes2011} who saw that the jamming transition is sharp when adjusting the density.

We found relations (Eqs.~\ref{eq:diff_exact} and~\ref{eq:v_c_exact}) between the diffusion constant, the persistence length and our order parameter defined in Eq.~\ref{eq:order}. The data collapse in Fig.~\ref{fig:analytic} proves that the assumptions we made to derive these relations are justified. Moreover, it shows that these relations hold independently of cluster size, providing a method to determine the values of the alignment and self-propulsion strength directly from experiments. Although our model is simple, it describes features found in biological swarms. Therefore, our results suggest that similar mechanisms based on local rules may be found in living systems, even if there are also more long-range (e.g. signaling-based) biological decision making processes present.

\begin{acknowledgments}
We would like to thank Silke Henkes and Gerard Barkema for helpful discussions.
\end{acknowledgments}

\bibliography{collectivedynamics11_arXiv}

\begin{thebibliography}{26}%
\makeatletter
\providecommand \@ifxundefined [1]{%
 \@ifx{#1\undefined}
}%
\providecommand \@ifnum [1]{%
 \ifnum #1\expandafter \@firstoftwo
 \else \expandafter \@secondoftwo
 \fi
}%
\providecommand \@ifx [1]{%
 \ifx #1\expandafter \@firstoftwo
 \else \expandafter \@secondoftwo
 \fi
}%
\providecommand \natexlab [1]{#1}%
\providecommand \enquote  [1]{``#1''}%
\providecommand \bibnamefont  [1]{#1}%
\providecommand \bibfnamefont [1]{#1}%
\providecommand \citenamefont [1]{#1}%
\providecommand \href@noop [0]{\@secondoftwo}%
\providecommand \href [0]{\begingroup \@sanitize@url \@href}%
\providecommand \@href[1]{\@@startlink{#1}\@@href}%
\providecommand \@@href[1]{\endgroup#1\@@endlink}%
\providecommand \@sanitize@url [0]{\catcode `\\12\catcode `\$12\catcode
  `\&12\catcode `\#12\catcode `\^12\catcode `\_12\catcode `\%12\relax}%
\providecommand \@@startlink[1]{}%
\providecommand \@@endlink[0]{}%
\providecommand \url  [0]{\begingroup\@sanitize@url \@url }%
\providecommand \@url [1]{\endgroup\@href {#1}{\urlprefix }}%
\providecommand \urlprefix  [0]{URL }%
\providecommand \Eprint [0]{\href }%
\providecommand \doibase [0]{http://dx.doi.org/}%
\providecommand \selectlanguage [0]{\@gobble}%
\providecommand \bibinfo  [0]{\@secondoftwo}%
\providecommand \bibfield  [0]{\@secondoftwo}%
\providecommand \translation [1]{[#1]}%
\providecommand \BibitemOpen [0]{}%
\providecommand \bibitemStop [0]{}%
\providecommand \bibitemNoStop [0]{.\EOS\space}%
\providecommand \EOS [0]{\spacefactor3000\relax}%
\providecommand \BibitemShut  [1]{\csname bibitem#1\endcsname}%
\let\auto@bib@innerbib\@empty
\bibitem [{\citenamefont {Shaw}(1978)}]{Shaw1978}%
  \BibitemOpen
  \bibfield  {author} {\bibinfo {author} {\bibfnamefont {E.}~\bibnamefont
  {Shaw}},\ }\href@noop {} {\bibfield  {journal} {\bibinfo  {journal} {Am.
  Sci.}\ }\textbf {\bibinfo {volume} {66}},\ \bibinfo {pages} {166} (\bibinfo
  {year} {1978})}\BibitemShut {NoStop}%
\bibitem [{\citenamefont {Calovi}\ \emph {et~al.}(2014)\citenamefont {Calovi},
  \citenamefont {Lopez}, \citenamefont {Ngo}, \citenamefont {Sire},
  \citenamefont {Chat\'{e}},\ and\ \citenamefont {Theraulaz}}]{Calovi2013}%
  \BibitemOpen
  \bibfield  {author} {\bibinfo {author} {\bibfnamefont {D.~S.}\ \bibnamefont
  {Calovi}}, \bibinfo {author} {\bibfnamefont {U.}~\bibnamefont {Lopez}},
  \bibinfo {author} {\bibfnamefont {S.}~\bibnamefont {Ngo}}, \bibinfo {author}
  {\bibfnamefont {C.}~\bibnamefont {Sire}}, \bibinfo {author} {\bibfnamefont
  {H.}~\bibnamefont {Chat\'{e}}}, \ and\ \bibinfo {author} {\bibfnamefont
  {G.}~\bibnamefont {Theraulaz}},\ }\href
  {http://stacks.iop.org/1367-2630/16/i=1/a=015026} {\bibfield  {journal}
  {\bibinfo  {journal} {New J. Phys.}\ }\textbf {\bibinfo {volume} {16}},\
  \bibinfo {pages} {015026} (\bibinfo {year} {2014})}\BibitemShut {NoStop}%
\bibitem [{\citenamefont {Kessler}\ and\ \citenamefont
  {Levine}(1993)}]{Kessler1993}%
  \BibitemOpen
  \bibfield  {author} {\bibinfo {author} {\bibfnamefont {D.~A.}\ \bibnamefont
  {Kessler}}\ and\ \bibinfo {author} {\bibfnamefont {H.}~\bibnamefont
  {Levine}},\ }\href {http://pre.aps.org/abstract/PRE/v48/i6/p4801\_1}
  {\bibfield  {journal} {\bibinfo  {journal} {Phys. Rev. E}\ }\textbf {\bibinfo
  {volume} {48}},\ \bibinfo {pages} {4801} (\bibinfo {year}
  {1993})}\BibitemShut {NoStop}%
\bibitem [{\citenamefont {Gr\'{e}goire}\ \emph {et~al.}(2003)\citenamefont
  {Gr\'{e}goire}, \citenamefont {Chat\'{e}},\ and\ \citenamefont
  {Tu}}]{Gregoire2003}%
  \BibitemOpen
  \bibfield  {author} {\bibinfo {author} {\bibfnamefont {G.}~\bibnamefont
  {Gr\'{e}goire}}, \bibinfo {author} {\bibfnamefont {H.}~\bibnamefont
  {Chat\'{e}}}, \ and\ \bibinfo {author} {\bibfnamefont {Y.}~\bibnamefont
  {Tu}},\ }\href {\doibase 10.1016/S0167-2789(03)00102-7} {\bibfield  {journal}
  {\bibinfo  {journal} {Phys. D}\ }\textbf {\bibinfo {volume} {181}},\ \bibinfo
  {pages} {157} (\bibinfo {year} {2003})}\BibitemShut {NoStop}%
\bibitem [{\citenamefont {Nagano}(1998)}]{Nagano1998}%
  \BibitemOpen
  \bibfield  {author} {\bibinfo {author} {\bibfnamefont {S.}~\bibnamefont
  {Nagano}},\ }\href {\doibase 10.1103/PhysRevLett.80.4826} {\bibfield
  {journal} {\bibinfo  {journal} {Phys. Rev. Lett.}\ }\textbf {\bibinfo
  {volume} {80}},\ \bibinfo {pages} {4826} (\bibinfo {year}
  {1998})}\BibitemShut {NoStop}%
\bibitem [{\citenamefont {Rappel}\ \emph {et~al.}(1999)\citenamefont {Rappel},
  \citenamefont {Nicol}, \citenamefont {Sarkissian}, \citenamefont {Levine},\
  and\ \citenamefont {Loomis}}]{Rappel1999}%
  \BibitemOpen
  \bibfield  {author} {\bibinfo {author} {\bibfnamefont {W.-J.}\ \bibnamefont
  {Rappel}}, \bibinfo {author} {\bibfnamefont {A.}~\bibnamefont {Nicol}},
  \bibinfo {author} {\bibfnamefont {A.}~\bibnamefont {Sarkissian}}, \bibinfo
  {author} {\bibfnamefont {H.}~\bibnamefont {Levine}}, \ and\ \bibinfo {author}
  {\bibfnamefont {W.~F.}\ \bibnamefont {Loomis}},\ }\href {\doibase
  10.1103/PhysRevLett.83.1247} {\bibfield  {journal} {\bibinfo  {journal}
  {Phys. Rev. Lett.}\ }\textbf {\bibinfo {volume} {83}},\ \bibinfo {pages}
  {1247} (\bibinfo {year} {1999})}\BibitemShut {NoStop}%
\bibitem [{\citenamefont {Butt}\ \emph {et~al.}(2010)\citenamefont {Butt},
  \citenamefont {Mufti}, \citenamefont {Humayun}, \citenamefont {Rosenthal},
  \citenamefont {Khan}, \citenamefont {Khan},\ and\ \citenamefont
  {Molloy}}]{Butt12022010}%
  \BibitemOpen
  \bibfield  {author} {\bibinfo {author} {\bibfnamefont {T.}~\bibnamefont
  {Butt}}, \bibinfo {author} {\bibfnamefont {T.}~\bibnamefont {Mufti}},
  \bibinfo {author} {\bibfnamefont {A.}~\bibnamefont {Humayun}}, \bibinfo
  {author} {\bibfnamefont {P.~B.}\ \bibnamefont {Rosenthal}}, \bibinfo {author}
  {\bibfnamefont {S.}~\bibnamefont {Khan}}, \bibinfo {author} {\bibfnamefont
  {S.}~\bibnamefont {Khan}}, \ and\ \bibinfo {author} {\bibfnamefont {J.~E.}\
  \bibnamefont {Molloy}},\ }\href {\doibase 10.1074/jbc.M109.044792} {\bibfield
   {journal} {\bibinfo  {journal} {J. Biol. Chem.}\ }\textbf {\bibinfo {volume}
  {285}},\ \bibinfo {pages} {4964} (\bibinfo {year} {2010})}\BibitemShut
  {NoStop}%
\bibitem [{\citenamefont {Vicsek}\ \emph {et~al.}(1995)\citenamefont {Vicsek},
  \citenamefont {Czir\'{o}k}, \citenamefont {Ben-Jacob}, \citenamefont
  {Cohen},\ and\ \citenamefont {Shochet}}]{Vicsek1995}%
  \BibitemOpen
  \bibfield  {author} {\bibinfo {author} {\bibfnamefont {T.}~\bibnamefont
  {Vicsek}}, \bibinfo {author} {\bibfnamefont {A.}~\bibnamefont {Czir\'{o}k}},
  \bibinfo {author} {\bibfnamefont {E.}~\bibnamefont {Ben-Jacob}}, \bibinfo
  {author} {\bibfnamefont {I.}~\bibnamefont {Cohen}}, \ and\ \bibinfo {author}
  {\bibfnamefont {O.}~\bibnamefont {Shochet}},\ }\href
  {http://prl.aps.org/abstract/PRL/v75/i6/p1226\_1} {\bibfield  {journal}
  {\bibinfo  {journal} {Phys. Rev. Lett.}\ }\textbf {\bibinfo {volume} {75}},\
  \bibinfo {pages} {1226} (\bibinfo {year} {1995})}\BibitemShut {NoStop}%
\bibitem [{\citenamefont {Vicsek}\ and\ \citenamefont
  {Zafeiris}(2012)}]{Vicsek2012}%
  \BibitemOpen
  \bibfield  {author} {\bibinfo {author} {\bibfnamefont {T.}~\bibnamefont
  {Vicsek}}\ and\ \bibinfo {author} {\bibfnamefont {A.}~\bibnamefont
  {Zafeiris}},\ }\href {\doibase 10.1016/j.physrep.2012.03.004} {\bibfield
  {journal} {\bibinfo  {journal} {Phys. Rep.}\ }\textbf {\bibinfo {volume}
  {517}},\ \bibinfo {pages} {71} (\bibinfo {year} {2012})}\BibitemShut
  {NoStop}%
\bibitem [{\citenamefont {Liu}\ and\ \citenamefont {Nagel}(1998)}]{Liu1998}%
  \BibitemOpen
  \bibfield  {author} {\bibinfo {author} {\bibfnamefont {A.~J.}\ \bibnamefont
  {Liu}}\ and\ \bibinfo {author} {\bibfnamefont {S.~R.}\ \bibnamefont
  {Nagel}},\ }\href {\doibase 10.1038/23819} {\bibfield  {journal} {\bibinfo
  {journal} {Nature}\ }\textbf {\bibinfo {volume} {396}},\ \bibinfo {pages}
  {21} (\bibinfo {year} {1998})}\BibitemShut {NoStop}%
\bibitem [{\citenamefont {Henkes}\ \emph {et~al.}(2011)\citenamefont {Henkes},
  \citenamefont {Fily},\ and\ \citenamefont {Marchetti}}]{Henkes2011}%
  \BibitemOpen
  \bibfield  {author} {\bibinfo {author} {\bibfnamefont {S.}~\bibnamefont
  {Henkes}}, \bibinfo {author} {\bibfnamefont {Y.}~\bibnamefont {Fily}}, \ and\
  \bibinfo {author} {\bibfnamefont {M.~C.}\ \bibnamefont {Marchetti}},\ }\href
  {\doibase 10.1103/PhysRevE.84.040301} {\bibfield  {journal} {\bibinfo
  {journal} {Phys. Rev. E}\ }\textbf {\bibinfo {volume} {84}},\ \bibinfo
  {pages} {040301} (\bibinfo {year} {2011})}\BibitemShut {NoStop}%
\bibitem [{\citenamefont {D'Orsogna}\ \emph {et~al.}(2006)\citenamefont
  {D'Orsogna}, \citenamefont {Chuang}, \citenamefont {Bertozzi},\ and\
  \citenamefont {Chayes}}]{D'Orsogna2006}%
  \BibitemOpen
  \bibfield  {author} {\bibinfo {author} {\bibfnamefont {M.~R.}\ \bibnamefont
  {D'Orsogna}}, \bibinfo {author} {\bibfnamefont {Y.~L.}\ \bibnamefont
  {Chuang}}, \bibinfo {author} {\bibfnamefont {A.~L.}\ \bibnamefont
  {Bertozzi}}, \ and\ \bibinfo {author} {\bibfnamefont {L.~S.}\ \bibnamefont
  {Chayes}},\ }\href {\doibase 10.1103/PhysRevLett.96.104302} {\bibfield
  {journal} {\bibinfo  {journal} {Phys. Rev. Lett.}\ }\textbf {\bibinfo
  {volume} {96}},\ \bibinfo {pages} {104302} (\bibinfo {year}
  {2006})}\BibitemShut {NoStop}%
\bibitem [{\citenamefont {Nguyen}\ \emph {et~al.}(2012)\citenamefont {Nguyen},
  \citenamefont {Jankowski},\ and\ \citenamefont {Glotzer}}]{Nguyen2012}%
  \BibitemOpen
  \bibfield  {author} {\bibinfo {author} {\bibfnamefont {N.~H.~P.}\
  \bibnamefont {Nguyen}}, \bibinfo {author} {\bibfnamefont {E.}~\bibnamefont
  {Jankowski}}, \ and\ \bibinfo {author} {\bibfnamefont {S.~C.}\ \bibnamefont
  {Glotzer}},\ }\href {\doibase 10.1103/PhysRevE.86.011136} {\bibfield
  {journal} {\bibinfo  {journal} {Phys. Rev. E}\ }\textbf {\bibinfo {volume}
  {86}},\ \bibinfo {pages} {011136} (\bibinfo {year} {2012})}\BibitemShut
  {NoStop}%
\bibitem [{\citenamefont {Wang}\ and\ \citenamefont {Kuspa}(1997)}]{Wang1997}%
  \BibitemOpen
  \bibfield  {author} {\bibinfo {author} {\bibfnamefont {B.}~\bibnamefont
  {Wang}}\ and\ \bibinfo {author} {\bibfnamefont {A.}~\bibnamefont {Kuspa}},\
  }\href {\doibase 10.1126/science.277.5323.251} {\bibfield  {journal}
  {\bibinfo  {journal} {Science}\ }\textbf {\bibinfo {volume} {277}},\ \bibinfo
  {pages} {251} (\bibinfo {year} {1997})}\BibitemShut {NoStop}%
\bibitem [{\citenamefont {Tunstr{\o}m}\ \emph {et~al.}(2013)\citenamefont
  {Tunstr{\o}m}, \citenamefont {Katz}, \citenamefont {Ioannou}, \citenamefont
  {Huepe}, \citenamefont {Lutz},\ and\ \citenamefont {Couzin}}]{Tunstrom2013}%
  \BibitemOpen
  \bibfield  {author} {\bibinfo {author} {\bibfnamefont {K.}~\bibnamefont
  {Tunstr{\o}m}}, \bibinfo {author} {\bibfnamefont {Y.}~\bibnamefont {Katz}},
  \bibinfo {author} {\bibfnamefont {C.~C.}\ \bibnamefont {Ioannou}}, \bibinfo
  {author} {\bibfnamefont {C.}~\bibnamefont {Huepe}}, \bibinfo {author}
  {\bibfnamefont {M.~J.}\ \bibnamefont {Lutz}}, \ and\ \bibinfo {author}
  {\bibfnamefont {I.~D.}\ \bibnamefont {Couzin}},\ }\href {\doibase
  10.1371/journal.pcbi.1002915} {\bibfield  {journal} {\bibinfo  {journal}
  {PLoS Comput. Biol.}\ }\textbf {\bibinfo {volume} {9}},\ \bibinfo {pages}
  {e1002915} (\bibinfo {year} {2013})}\BibitemShut {NoStop}%
\bibitem [{\citenamefont {McCann}\ \emph {et~al.}(2010)\citenamefont {McCann},
  \citenamefont {Kriebel}, \citenamefont {Parent},\ and\ \citenamefont
  {Losert}}]{McCann2010}%
  \BibitemOpen
  \bibfield  {author} {\bibinfo {author} {\bibfnamefont {C.~P.}\ \bibnamefont
  {McCann}}, \bibinfo {author} {\bibfnamefont {P.~W.}\ \bibnamefont {Kriebel}},
  \bibinfo {author} {\bibfnamefont {C.~A.}\ \bibnamefont {Parent}}, \ and\
  \bibinfo {author} {\bibfnamefont {W.}~\bibnamefont {Losert}},\ }\href
  {\doibase 10.1242/jcs.060137} {\bibfield  {journal} {\bibinfo  {journal} {J.
  Cell Sci.}\ }\textbf {\bibinfo {volume} {123}},\ \bibinfo {pages} {1724}
  (\bibinfo {year} {2010})}\BibitemShut {NoStop}%
\bibitem [{\citenamefont {Vasiev}\ \emph {et~al.}(1997)\citenamefont {Vasiev},
  \citenamefont {Siegert},\ and\ \citenamefont {Weijer}}]{Vasiev1997}%
  \BibitemOpen
  \bibfield  {author} {\bibinfo {author} {\bibfnamefont {B.}~\bibnamefont
  {Vasiev}}, \bibinfo {author} {\bibfnamefont {F.}~\bibnamefont {Siegert}}, \
  and\ \bibinfo {author} {\bibfnamefont {C.~J.}\ \bibnamefont {Weijer}},\
  }\href {\doibase 10.1006/jtbi.1996.0282} {\bibfield  {journal} {\bibinfo
  {journal} {J. Theor. Biol.}\ }\textbf {\bibinfo {volume} {184}},\ \bibinfo
  {pages} {441} (\bibinfo {year} {1997})}\BibitemShut {NoStop}%
\bibitem [{\citenamefont {Saragosti}\ \emph {et~al.}(2011)\citenamefont
  {Saragosti}, \citenamefont {Calvez}, \citenamefont {Bournaveas},
  \citenamefont {Perthame}, \citenamefont {Buguin},\ and\ \citenamefont
  {Silberzan}}]{Saragosti2011}%
  \BibitemOpen
  \bibfield  {author} {\bibinfo {author} {\bibfnamefont {J.}~\bibnamefont
  {Saragosti}}, \bibinfo {author} {\bibfnamefont {V.}~\bibnamefont {Calvez}},
  \bibinfo {author} {\bibfnamefont {N.}~\bibnamefont {Bournaveas}}, \bibinfo
  {author} {\bibfnamefont {B.}~\bibnamefont {Perthame}}, \bibinfo {author}
  {\bibfnamefont {A.}~\bibnamefont {Buguin}}, \ and\ \bibinfo {author}
  {\bibfnamefont {P.}~\bibnamefont {Silberzan}},\ }\href {\doibase
  10.1073/pnas.1101996108} {\bibfield  {journal} {\bibinfo  {journal} {Proc.
  Natl. Acad. Sci. USA}\ }\textbf {\bibinfo {volume} {108}},\ \bibinfo {pages}
  {16235} (\bibinfo {year} {2011})}\BibitemShut {NoStop}%
\bibitem [{\citenamefont {Czir\'{o}k}\ \emph {et~al.}(1996)\citenamefont
  {Czir\'{o}k}, \citenamefont {Ben-Jacob}, \citenamefont {Cohen},\ and\
  \citenamefont {Vicsek}}]{Czirok1996}%
  \BibitemOpen
  \bibfield  {author} {\bibinfo {author} {\bibfnamefont {A.}~\bibnamefont
  {Czir\'{o}k}}, \bibinfo {author} {\bibfnamefont {E.}~\bibnamefont
  {Ben-Jacob}}, \bibinfo {author} {\bibfnamefont {I.}~\bibnamefont {Cohen}}, \
  and\ \bibinfo {author} {\bibfnamefont {T.}~\bibnamefont {Vicsek}},\ }\href
  {http://www.ncbi.nlm.nih.gov/pubmed/9965259} {\bibfield  {journal} {\bibinfo
  {journal} {Phys. Rev. E}\ }\textbf {\bibinfo {volume} {54}},\ \bibinfo
  {pages} {1791} (\bibinfo {year} {1996})}\BibitemShut {NoStop}%
\bibitem [{\citenamefont {Deforet}\ \emph {et~al.}(2014)\citenamefont
  {Deforet}, \citenamefont {van Ditmarsch}, \citenamefont {Carmona-Fontaine},\
  and\ \citenamefont {Xavier}}]{Deforet2014}%
  \BibitemOpen
  \bibfield  {author} {\bibinfo {author} {\bibfnamefont {M.}~\bibnamefont
  {Deforet}}, \bibinfo {author} {\bibfnamefont {D.}~\bibnamefont {van
  Ditmarsch}}, \bibinfo {author} {\bibfnamefont {C.}~\bibnamefont
  {Carmona-Fontaine}}, \ and\ \bibinfo {author} {\bibfnamefont {J.~B.}\
  \bibnamefont {Xavier}},\ }\href {\doibase 10.1039/c3sm53127a} {\bibfield
  {journal} {\bibinfo  {journal} {Soft Matter}\ }\textbf {\bibinfo {volume}
  {10}},\ \bibinfo {pages} {2405} (\bibinfo {year} {2014})}\BibitemShut
  {NoStop}%
\bibitem [{\citenamefont {Weijer}(2004)}]{Weijer2004}%
  \BibitemOpen
  \bibfield  {author} {\bibinfo {author} {\bibfnamefont {C.~J.}\ \bibnamefont
  {Weijer}},\ }\href {\doibase 10.1016/j.gde.2004.06.006} {\bibfield  {journal}
  {\bibinfo  {journal} {Curr. Opin. Genetics Dev.}\ }\textbf {\bibinfo {volume}
  {14}},\ \bibinfo {pages} {392} (\bibinfo {year} {2004})}\BibitemShut
  {NoStop}%
\bibitem [{\citenamefont {Landau}\ and\ \citenamefont
  {Lifshitz}(1987)}]{Landau1987}%
  \BibitemOpen
  \bibfield  {author} {\bibinfo {author} {\bibfnamefont {L.~D.}\ \bibnamefont
  {Landau}}\ and\ \bibinfo {author} {\bibfnamefont {E.~M.}\ \bibnamefont
  {Lifshitz}},\ }\href
  {http://scholar.google.com/scholar?hl=en\&btnG=Search\&q=intitle:Fluid+Mechanics+1987\#0}
  {\emph {\bibinfo {title} {{Fluid Mechanics}}}}\ (\bibinfo  {publisher}
  {{Butterworth-Heinemann}},\ \bibinfo {address} {{Oxford}},\ \bibinfo {year}
  {1987})\BibitemShut {NoStop}%
\bibitem [{Note1()}]{Note1}%
  \BibitemOpen
  \bibinfo {note} {We need to set $\chi $ because Stokes' paradox does not
  allow us to relate the translational and rotational viscosities $\eta $ and
  $\eta _R$ ~\cite {Landau1987}.}\BibitemShut {Stop}%
\bibitem [{Note2()}]{Note2}%
  \BibitemOpen
  \bibinfo {note} {See Supplemental Material at \texttt{\href{http://homepage.tudelft.nl/5u8n5/SupplMat.html}{http://homepage. tudelft.nl/5u8n5/SupplMat.html}} for movies of the various types of behavior displayed by the
  cluster}\BibitemShut {NoStop}%
\bibitem [{\citenamefont {Visscher}(1973)}]{Visscher1973}%
  \BibitemOpen
  \bibfield  {author} {\bibinfo {author} {\bibfnamefont {W.~M.}\ \bibnamefont
  {Visscher}},\ }\href
  {http://journals.aps.org/pra/abstract/10.1103/PhysRevA.7.1439} {\bibfield
  {journal} {\bibinfo  {journal} {Phys. Rev. A}\ }\textbf {\bibinfo {volume}
  {7}},\ \bibinfo {pages} {1439} (\bibinfo {year} {1973})}\BibitemShut {NoStop}%
\bibitem [{\citenamefont {Shapiro}(1988)}]{Shapiro1988}%
  \BibitemOpen
  \bibfield  {author} {\bibinfo {author} {\bibfnamefont {J.}~\bibnamefont
  {Shapiro}},\ }\href {\doibase 10.1038/scientificamerican0688-82} {\bibfield
  {journal} {\bibinfo  {journal} {Sci. Am.}\ }\textbf {\bibinfo {volume}
  {256}},\ \bibinfo {pages} {82} (\bibinfo {year} {1988})}\BibitemShut
  {NoStop}%
\end{thebibliography}%

\end{document}